\documentclass[a4paper]{jpconf}
\usepackage{cite} 
\usepackage[colorlinks=true, allcolors=blue]{hyperref}
\usepackage{graphicx}
\begin{document}
\title{Laser-cooling Cadmium Bosons and Fermions with \\Near Ultraviolet Triplet Excitations} 

\author{Kurt Gibble}
\address{Department of Physics, The Pennsylvania State University, University Park, PA, 16802 USA}
\ead{kgibble@psu.edu}

\begin{abstract}
Cadmium is laser-cooled and trapped with 
excitations to triplet states with UVA light, first using only the $67\,$kHz wide $326\,$nm intercombination line and subsequently, for large loading rates, the $25\,$MHz wide $361\,$nm $\,^3P_2$$\,\rightarrow\,$$\,^3D_3$ transition. Eschewing the hard UV $229\,$nm $^1S_0$$\,\rightarrow$$\,^1P_1$ 
transition, only small magnetic fields gradients, less than $6\,$G$\,$cm$^{-1}$, are required enabling 
a 100\% transfer of atoms from the $361\,$nm trap to the $326\,$nm narrow-line trap. 
All 8 stable cadmium isotopes are straightforwardly trapped, including two nuclear-spin-½ fermions that require no additional repumping. 
We observe evidence of $^3P_2$ collisions limiting the number of trapped metastable atoms, report isotope shifts for $^{111}$Cd and $^{113}$Cd of the $326\,$nm $^1S_0$$\,\rightarrow$$\,^3P_1$, $480\,$nm $^3P_1$$\,\rightarrow$$\,^3S_1$, and $361\,$nm $^3P_2$$\,\rightarrow$$\,^3D_3$ transitions, and measure the $^{114}$Cd 5s5p$\,^3P_2$$\,\rightarrow\,$5s5d$\,^3D_3$ transition frequency to be $830\,096\,573(15)\,$MHz.
\end{abstract}

\section{Introduction}
Strontium, ytterbium and other atoms with two valence electrons are attractive for state-of-the-art quantum sensors such as optical-lattice atomic clocks and atom interferometers \cite{ClockRevRMP15, BACON2021_AlSrYb, BizeMore254,Ohmae2020_HgYbSrloop,TinoSr8788EEP2014, MAGIS100_2021,Tino_AIrev_2021, Poli_CdSr_Source2023}, for quantum gas experiments \cite{Takahashi_YbBEC2003, PTB_Ca_BEC2009, Schreck_Sr84BEC_2009, Killian_Sr84BEC_2009, Rey_Fermigases_SU(N)_2014, Chen2022CWBEC}, 
and for quantum computation, information and simulation \cite{SU_N_Group2_2010, Daley_Group2_QC_QI_2011, Foss_Feig_QI_2019, AE_Rydberg_Gate2020, Bloch2012QSim, QSim_Takahashi2020}.
Yb, Ca, Cd and Hg have more than three stable spin 0 isotopes, enabling tests of fundamental physics through precise measurements of isotope shifts (ISs)  \cite{2020-Yb,2021-Yb, 2021-Yb2, 2022-Yb, CaISS, KolkowitzSrIS, CdISS}. The broad applicability of alkaline-earth-like atoms to high-coherence measurements is facilitated by their nearly forbidden clock and other intercombination transitions, from the singlet $^1S_0$ ground state to triplet $^3P_{J=0,1,2}$ states, as well as fast laser cooling, on the broad electric dipole allowed $^1S_0$$\,\rightarrow$$\,^1P_1$ transition for most atoms (Ca, Sr, Yb, Mg, Cd and Zn). 
These atoms have clock transitions with mHz natural linewidths, their $^1S_0$$\,\rightarrow$$\,^3P_1$ intercombination transitions have sub-MHz linewidths, which allow laser cooling to $\mu$K temperatures and lower, and their paired electron spins provide a natural insensitivity to perturbations from external magnetic fields. For the currently most accurate optical lattice clocks, based on Sr and Yb, one of the leading systematic errors is the frequency shift due to blackbody radiation (BBR) \cite{ClockRevRMP15, BACON2021_AlSrYb, Ohmae2020_HgYbSrloop}.  
An attractive feature of Cd and Hg optical lattice clocks is their order of magnitude smaller sensitivities of their clock transitions to BBR than the Sr and Yb clock transitions \cite{CdMag,Derevianko_ZnCdBBR_2019,BizeMore254,Ohmae2020_HgYbSrloop}.

Previous laser cooling of Cd and Hg has used hard UV laser light, exciting the broad $91\,$MHz wide $229\,$nm $^1S_0$$\,\rightarrow$$\,^1P_1$ Cd transition \cite{MonroeCd,Katori229,CdMag} and the $1.3\,$MHz wide $254\,$nm $^1S_0$$\,\rightarrow$$\,^3P_1$ Hg intercombination transition \cite{BizeMore254,Ohmae2020_HgYbSrloop,Stellmer_Hg_highphasespace2022}. 
These UVC sources are subject to degradation of non-linear crystals and optical coatings, making long-term operation, especially autonomous operation, challenging \cite{BizeMore254,Stellmer_Hg_highphasespace2022}. Here, we demonstrate large loading rates and efficient laser-cooling of Cd using only near UV excitations to triplet states, with $326\,$nm and $361\,$nm light, in addition to two low-power blue optical pumping lasers, importantly, without having used $229\,$nm $^1S_0$$\,\rightarrow$$\,^1P_1$ UVC light \cite{MonroeCd, Katori229, Poli229, TruppeFermionISS}. 
The multilevel lower state structure of the $361\,$nm broad-line magneto-optic trap (MOT) does not require the usual large magnetic field gradients of $^1S_0$$\,\rightarrow$$\,^1P_1$ MOTs and provides polarization gradient cooling, facilitating a 100\% transfer of atoms to the $326\,$nm narrow-line MOT.

\begin{figure}
\begin{center}
\includegraphics[width=4in]{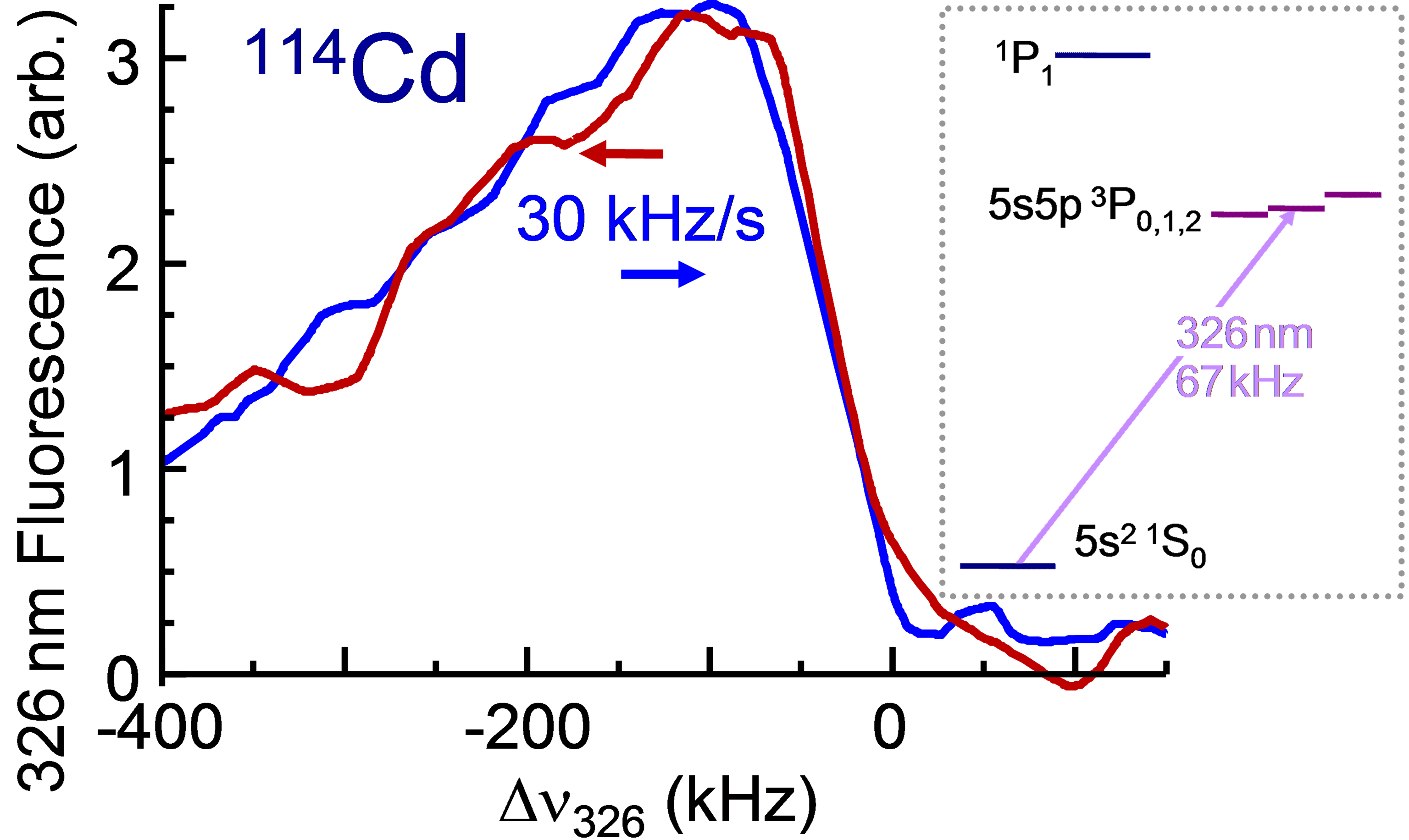}
\end{center}
\caption{\label{fig:326MOT}
Fluorescence of a Cd MOT that uses only the $67\,$kHz wide $^1S_0$$\,\rightarrow$$\,^3P_1$ $326\,$nm intercombination transition (see inset) to capture and trap atoms from an effusive source at $133\,$C. 
The blue (red) curve is the average of four frequency scans at $+(-)30\,$kHz$\,$s$^{-1}$ and shows a sharp blue edge near the $^1S_0$$\,\rightarrow$$\,^3P_1$ resonance.}
\end{figure}

We first demonstrate laser-cooling and trapping of approximately $10^4$ Cd atoms from an effusive source using only the $67\,$kHz wide $326\,$nm $^1S_0$$\,\rightarrow$ $\,^3P_1$ intercombination transition \cite{Cd3P1Hyperfine1964, Cd3p1Lifetime1989} (see Fig.~\ref{fig:326MOT}). 
While the narrow transition linewidth enables low Doppler cooling temperatures \cite{CdMag}, in the absence of stimulated processes \cite{ThompsonNarrowStimCooling}, the slow $^3P_1$ spontaneous emission limits the rate at which $326\,$nm photons can slow Cd atoms as they cross the laser beams. 
This is the narrowest transition used to capture atoms from a room-temperature (or hotter) source into an atom trap \cite{Vladen_3P1only_and1P12015,YbSingletslowerTripletMOT,YbCoreMOT}. 

\begin{figure*}
\begin{center}
\includegraphics[width=6.3in]
{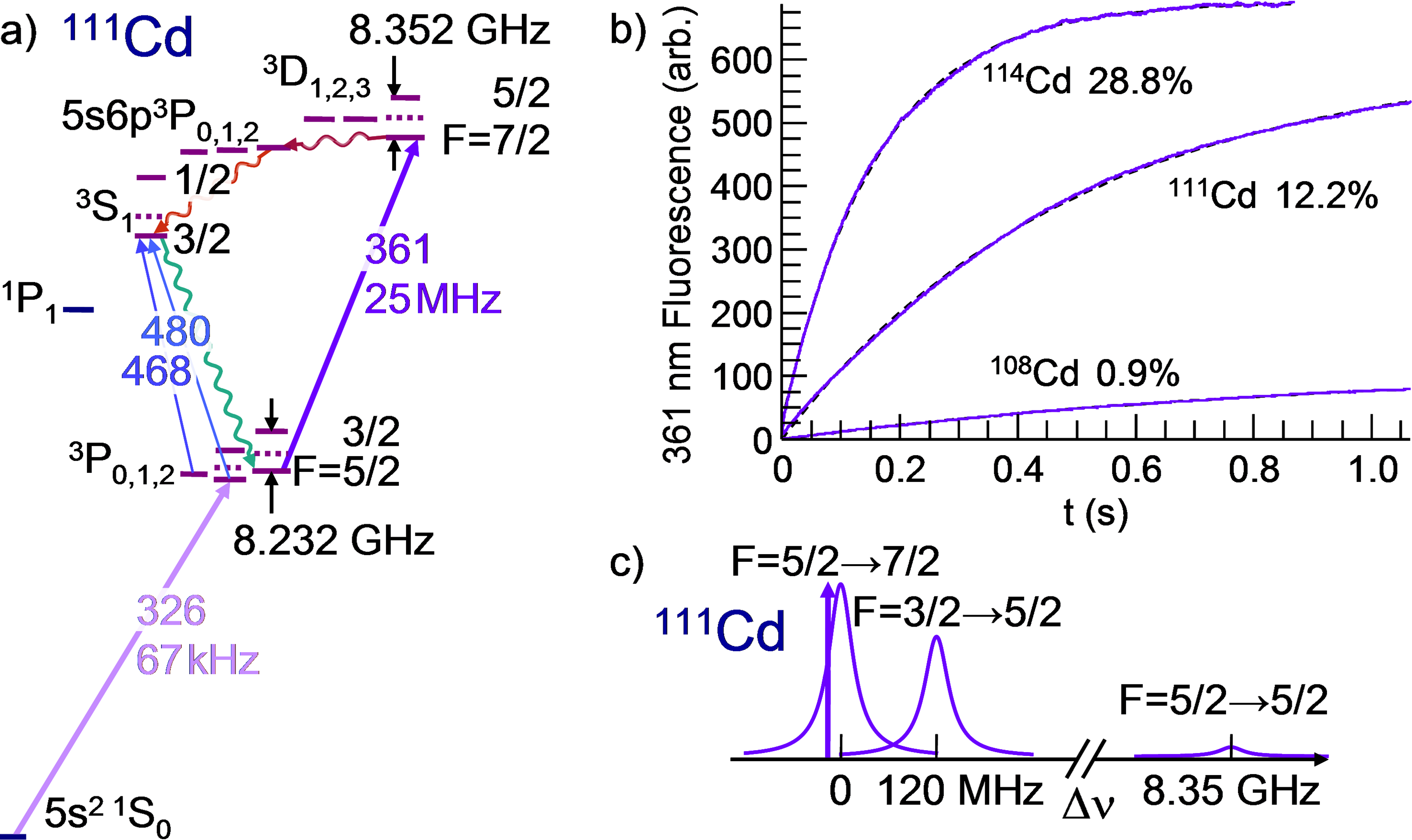}\caption{\label{fig:361MOT}
a) Energy levels of Cd fermions (solid) and bosons (dotted), where the annotated transition wavelengths are in nanometers. Atoms are excited to $^3P_1$ and optically pumped to $^3P_2$ via  $^3S_1$ with $480\,$nm and $468\,$nm light. 
The $361\,$nm $\,^3P_2$$\,\rightarrow\,$$\,^3D_3$ MOT transition is nearly closed, with a slow $11.4\,\mu$m and $1.4\,\mu$m cascade to 5s6p$\,^3P_2$ and $^3S_1$. The fermion hyperfine splittings from Table~\ref{tab:HFS} are not depicted to scale and those of the $^1P_1$, $^3D_{1,2}$ and 5s6p $^3P_{1,2}$ states are omitted. 
b) Exciting the $25\,$MHz wide $361\,$nm $\,^3P_2$$\,\rightarrow\,$$\,^3D_3$ transition efficiently captures the 8 Cd bosonic and fermionic isotopes. The faster saturation of the $^{114}$Cd and $^{111}$Cd loading than that for the $0.9\%$ abundant $^{108}$Cd suggests that $\,^3P_2$ collisional loss limits the number of atoms trapped.
c) To trap $^{111}$Cd and $^{113}$Cd, the $361\,$nm laser is red-detuned from the $^3P_2$ $F=\frac{5}{2}\,\rightarrow\,$$^3D_3\,F=\frac{7}{2}$ MOT transition.  The inverted $^{111}$Cd and $^{113}$Cd fermion hyperfine structure and the larger hyperfine splitting of $^3D_3$ than that of $^3P_2$ in Table \ref{tab:HFS} enable convenient repumping (and cooling) of $^3P_2\,F=\frac{3}{2}$ atoms via the $^3P_2\,F=\frac{3}{2}\,\rightarrow\,^3D_3\,F=\frac{5}{2}$ transition, from which the $361\,$nm laser is further red-detuned by $120\,$MHz for $^{111}$Cd, and $125\,$MHz for $^{113}$Cd.
}
\end{center}
\end{figure*}

\begin{figure}
\begin{center}
\includegraphics[width=3.7in]{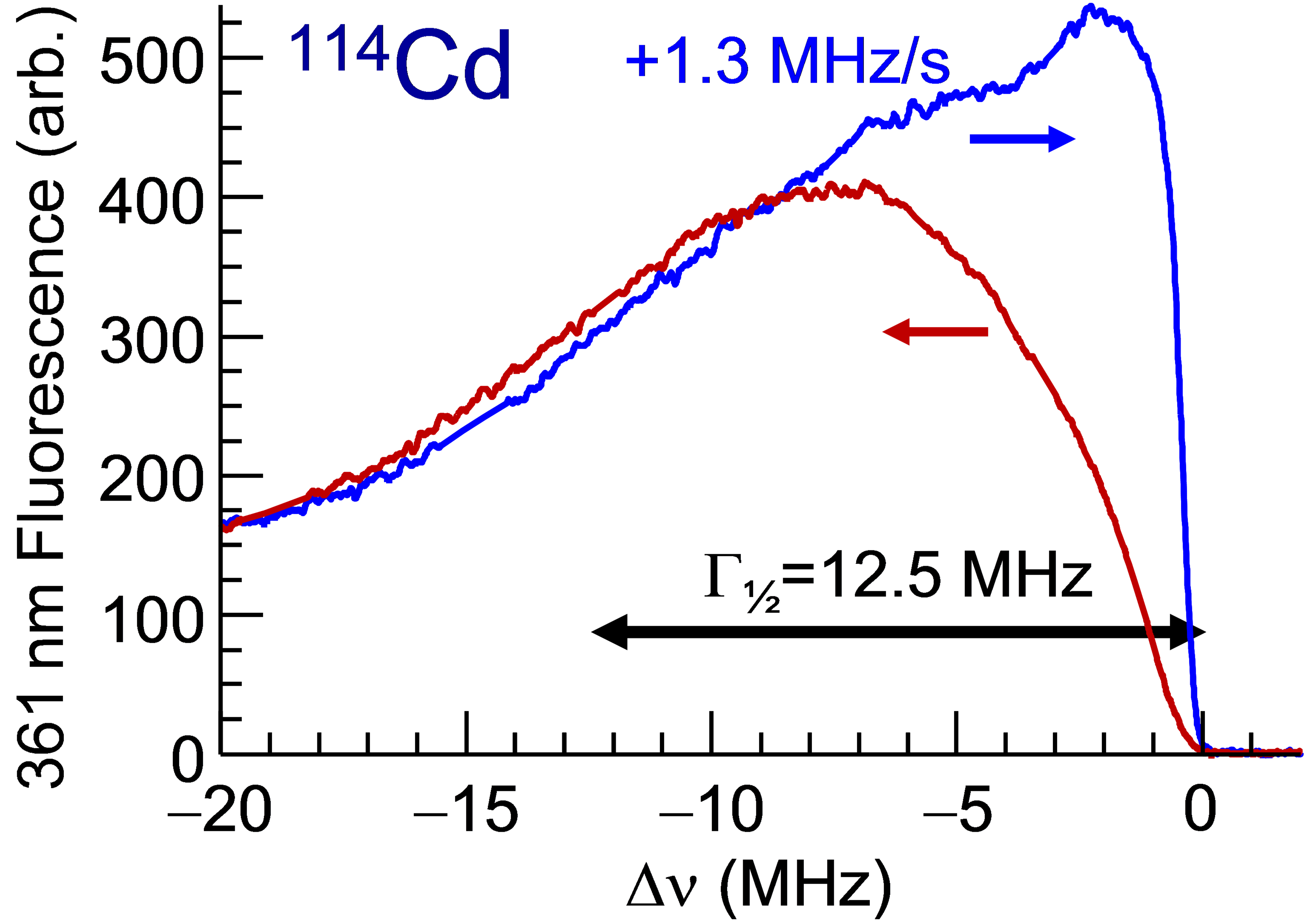}
\caption{\label{fig:361MOTscan}
$361\,$nm metastable MOT fluorescence as the laser frequency is tuned across the $^3P_2$$\,\rightarrow$$\,^3D_3$ resonance.} 
\end{center}
\end{figure}

Next, we dramatically increase the laser-cooling capture rate using the dipole-allowed $^3P_2$$\,\rightarrow$$\,^3D_3$ transition. For Cd, this transition is in the near UV, $361\,$nm in  Fig.~\ref{fig:361MOT}a), and has a reasonably broad $25\,$MHz linewidth. 
With other alkaline-earth atoms, this transition has been used in conjunction with the singlet transition to cool atoms to increase MOT and lattice loading fractions \cite{Ca3p23D3MOT,Rasel3D3Det,KatoriSr3P2Cool} and for high signal-to-noise detection \cite{Rasel3D3Det}. Excitations from $^3P_J$ to other states, such as $^3P_0$$\,\rightarrow$$\,^3D_1$, have been used to cool to sub-recoil temperatures \cite{LudlowDarkCool,Ludlow3D1Latt}; 
and $^3P_{0,1}$$\,\rightarrow$$\,^3S_1$, to increase the number of Sr atoms loaded into a magnetic trap, and subsequently an intercombination MOT \cite{GretchenCool3P2}; 
and $^3P_1$$\,\rightarrow$$\,^1D_2$ and $^3P_1$$\,\rightarrow$$\,^1S_0$, to increase the scattering rate for a Ca intercombination MOT \cite{CaQuenched,PTBNISTCa}. 

Exciting the $361\,$nm UVA transition, we efficiently capture room temperature atoms and load a metastable $^3P_2$$\,\rightarrow$$\,^3D_3$ MOT with all 6 of the stable nuclear-spin-zero Cd bosonic isotopes, $^{106}$Cd to $^{116}$Cd. 
We further show that no additional lasers are needed for hyperfine pumping to efficiently capture both nuclear-spin $I=\frac{1}{2}$ fermionic isotopes, $^{111}$Cd and $^{113}$Cd, in the metastable MOT. We observe evidence of $^3P_2$ collisional loss, report the ISs of three transitions, 
the $326\,$nm $^1S_0$$\,\rightarrow$$\,^3P_1$, 
the $480\,$nm $^3P_1$$\,\rightarrow$$\,^3S_1$ and 
the $361\,$nm $^3P_2$$\,\rightarrow$$\,^3D_3$ transition, and measure the absolute frequency of the $361\,$nm $^3P_2$$\,\rightarrow$$\,^3D_3$ transition, improving its previous uncertainty by a factor of 20. This $^3P_2$$\,\rightarrow$$\,^3D_3$ trapping may be applied to other alkaline-earth-like atoms, for example, Hg, where it can be challenging to maintain sufficient $254\,$nm power to capture many atoms in a  $^1S_0$$\,\rightarrow$$\,^3P_1$ MOT \cite{BizeMore254,Stellmer_Hg_highphasespace2022}.

\begin{center}
\begin{table}[h]
\caption{\label{tab:HFS}
$^{111}$Cd, and $^{113}$Cd, hyperfine splittings of Fig.~\ref{fig:361MOT}a) in MHz \cite{Cd3S1HyperfineEPJ, Cd3S1HyperfineRoyal,Cd3P1Hyperfine1964, LichtenCd3P2Hyperfine,Cd3DHyperfine}.}
\centering
\begin{tabular}{@{}*{7}{l}}
\br
A & $^3S_1$ & $^3P_1$ & $^3P_2$ & $^3D_3$\\
\mr
111    &  $-11626.2(13)$ &  $-6185.72(2)$ & $-8232.341(2)$ & $-8352.04(9)$ \\
113    &  $-12162.9(14)$  &  $-6470.79(2)$ & $-8611.586(4)$ & $-8736.88(9)$ \\
\br
\end{tabular}
\end{table}
\end{center}

\begin{center}
\begin{table}[h]
\caption{\label{tab:ISS}
Isotope shifts of Cd I fermions relative to $^{114}$Cd, in MHz. We measure the $326\,$nm $^1S_0\,\rightarrow \,^3P_1\,F=\frac{3}{2}$, the $361\,$nm $^3P_2\,F=\frac{5}{2} \rightarrow \,^3D_3\,F=\frac{7}{2}$ and the $480\,$nm $^3P_1\,F=\frac{3}{2} \rightarrow \,^3S_1 \, F=\frac{3}{2}$ hyperfine transitions and calculate the transition centers-of-mass using the $^3P_1$, $^3P_2$, $^3S_1$ and $^3D_3$ hyperfine splittings in Table \ref{tab:HFS}.}
\centering
\begin{tabular}{@{}*{7}{l}}
\br
A & $326\,$nm & $361\,$nm & $480\,$nm\\ 
\mr
111    &  $861.6(10)$ &  $-227.4(10)$ & $-334.2(11)$ \\
113    &  $370.8(10)$  &  $ -75.3(10)$ & $-135.0(11)$ \\
\br
\end{tabular}
\end{table}
\end{center}

\section{Cadmium Intercombination and Metastable MOT's}
\subsection{Narrow-line $326\,$nm intercombination MOT}
Trapping a detectable number of Cd atoms using only the $67\,$kHz wide $326\,$nm intercombination line in Fig.~\ref{fig:326MOT} is an important first step to bootstrap to the large loading rates of the $361\,$nm $^3P_2$$\,\rightarrow$$\,^3D_3$ MOT. Our effusive source of Cd atoms with natural isotopic abundances has a $1.2\,$cm diameter, is $2.2\,$cm from the center of the MOT, and is heated to $120\,$C to $142\,$C. 
We begin with $50$ to $150\,$mW of $326\,$nm light that is split into three retro-reflected beams with $e^{-2}$ diameters of $8.5\,$mm to form a MOT with a quadrupole field gradient of $0.5\,$G$\,$cm$^{-1}$ \cite{Cd326Details}. 
To increase the velocity capture range, an acousto-optic modulator (AOM) frequency modulates (FM) the $326\,$nm light at $50\,$kHz with a peak-to peak amplitude of $8.6\,$MHz \cite{CdMag}. 
We trap the six stable bosonic and fermionic isotopes, $^{110}$Cd to $^{116}$Cd, which have abundances greater than $7.5\%$. 
The number of trapped $^{114}$Cd atoms versus frequency in Fig.~\ref{fig:326MOT} shows a sharp blue edge \cite{RolstenBlue}, which we use to measure isotope shifts \cite{CdISS}. 
The $^{111}$Cd and $^{113}$Cd fermions are trapped using their $^1S_0\,F=\frac{1}{2}\,\rightarrow$$\,^3P_1\,F=\frac{3}{2}$ hyperfine components and their ISs, given in Table~\ref{tab:ISS}, are consistent and more precise, as are the bosonic ISs \cite{CdISS}, with other recent measurements \cite{TruppeFermionISS}. To trap $^{106}$Cd and $^{108}$Cd, which have $1.25\%$ and $0.89\%$ abundances, we use the $361\,$nm $^3P_2$$\,\rightarrow$$\,^3D_3$ MOT, described next, to increase the loading rate.

\subsection{$361\,$nm Metastable MOT}
With cold atoms trapped in the $326\,$nm intercombination MOT, it is straightforward to successively add and frequency tune the lasers for the $480\,$nm $^3P_1$$\,\rightarrow$$\,^3S_1$ optical pumping, the $361\,$nm $^3P_2$$\,\rightarrow$$\,^3D_3$ metastable MOT, and the $468\,$nm $^3P_0$$\,\rightarrow$$\,^3S_1$ optical pumping [see Fig.~\ref{fig:361MOT}a)]. 
The optical pumping to $^3P_2$ is well saturated with $100\,\mu$W of power in $12.6\,$mm diameter $468\,$nm and $8.7\,$mm diameter $480\,$nm laser beams. The $480\,$nm $^3P_1$$\,\rightarrow$$\,^3S_1$ resonance can be easily found as this pumping extinguishes the $326\,$nm MOT fluorescence. We use $50$ to $200\,$mW of $361\,$nm light that is intensity controlled and frequency shifted by an AOM before splitting into three retro-reflected MOT beams with $e^{-2}$ diameters of $10.4\,$mm. The number of atoms depends weakly on the optimal MOT quadrupole gradient of $5.7\,$G$\,$cm$^{-1}$, notably smaller than the of order $200\,$G$\,$cm$^{-1}$ gradients of $^1S_0$$\,\rightarrow$$\,^1P_1$ singlet MOT’s \cite{MonroeCd,Katori229,CdMag,Vladen_3P1only_and1P12015}. 
Here, the $326\,$nm FM depth is larger, $16.3\,$MHz, consistent with the  larger capture velocity of the $25\,$MHz wide $361\,$nm $^3P_2$$\,\rightarrow$$\,^3D_3$ MOT \cite{MonsterMOT}. From the saturation of the $361\,$nm fluorescence versus a pulsed intensity of the MOT beams, the number of trapped $^{114}$Cd atoms is greater than $6\times 10^6$. We trap all 6 spin 0 bosonic isotopes, which have no hyperfine structure [dashed levels in Fig.~\ref{fig:361MOT}a)], including $^{106}$Cd and $^{108}$Cd with $1\%$ abundances [Fig.~\ref{fig:361MOT}b)]. As discussed below, the loading behavior in Fig.~\ref{fig:361MOTscan} for increasing frequency sweeps (blue curve), as well as the different loading time constants in Fig.~\ref{fig:361MOT}b), suggests that $^3P_2$ collisional loss limits the number of trapped metastable atoms.

In addition to the 6 Cd bosonic isotopes, the $361\,$nm $^3P_2$$\,\rightarrow$$\,^3D_3$ metastable MOT efficiently captures $^{111}$Cd [Fig.~\ref{fig:361MOT}b)] and $^{113}$Cd nuclear-spin-½ fermions. Whereas additional repumping lasers are normally required for $J\,\rightarrow\,J+1$ MOT transitions with $J,I \ne 0$, especially for large nuclear spins \cite{KatoriSr3P2Cool,GretchenCool3P2}, no additional laser frequencies are required here for Cd. This is enabled by the negative nuclear magnetic moments of $^{111}$Cd [Fig.~\ref{fig:361MOT}a)] and $^{113}$Cd and their $I=\frac{1}{2}$ nuclear spins. 
Each $J \ne 0$ state has only two hyperfine components, $F=J \pm I$, from the nuclear magnetic dipole interaction, with no electric quadrupole shifts for $I\le \frac{1}{2}$. 
Table \ref{tab:HFS} gives the Cd hyperfine splittings and we highlight that the $^3D_3$ hyperfine splitting for $^{111}$Cd($^{113}$Cd) is $120(125)\,$MHz larger than the $^3P_2$ hyperfine splitting \cite{LichtenCd3P2Hyperfine,Cd3DHyperfine}. Combined with a negative nuclear magnetic moment, which yields the “inverted” ($^3S_1$, $^3P_1$,) $^3P_2$ and $^3D_3$ hyperfine structure, the $^3P_2\,F=\frac{3}{2}\,\rightarrow\,$$^3D_3\,F=\frac{5}{2}$ transition in Fig.~\ref{fig:361MOT}c) is therefore $120\,$MHz blue detuned from the $^3P_2\,F=\frac{5}{2}\,\rightarrow\,^3D_3 \, F=\frac{7}{2}$ metastable MOT cycling transition. As a result, atoms lost from the cycling transition to $^3P_2\,F=\frac{3}{2}$ are naturally repumped, cooled and trapped via the $^3P_2\,F=\frac{3}{2},\rightarrow\,$$^3D_3\,F=\frac{5}{2}$ transition by the red detuned $361\,$nm MOT beams. 
The nuclear magnetic moment and resulting hyperfine splittings of $^{113}$Cd in Table \ref{tab:HFS} are the same, within $5\%$, as for $^{111}$Cd, allowing similarly large loading rates of $^{113}$Cd as in Fig.~\ref{fig:361MOT}b).

\begin{figure}
\begin{center}
\includegraphics[width=4in]{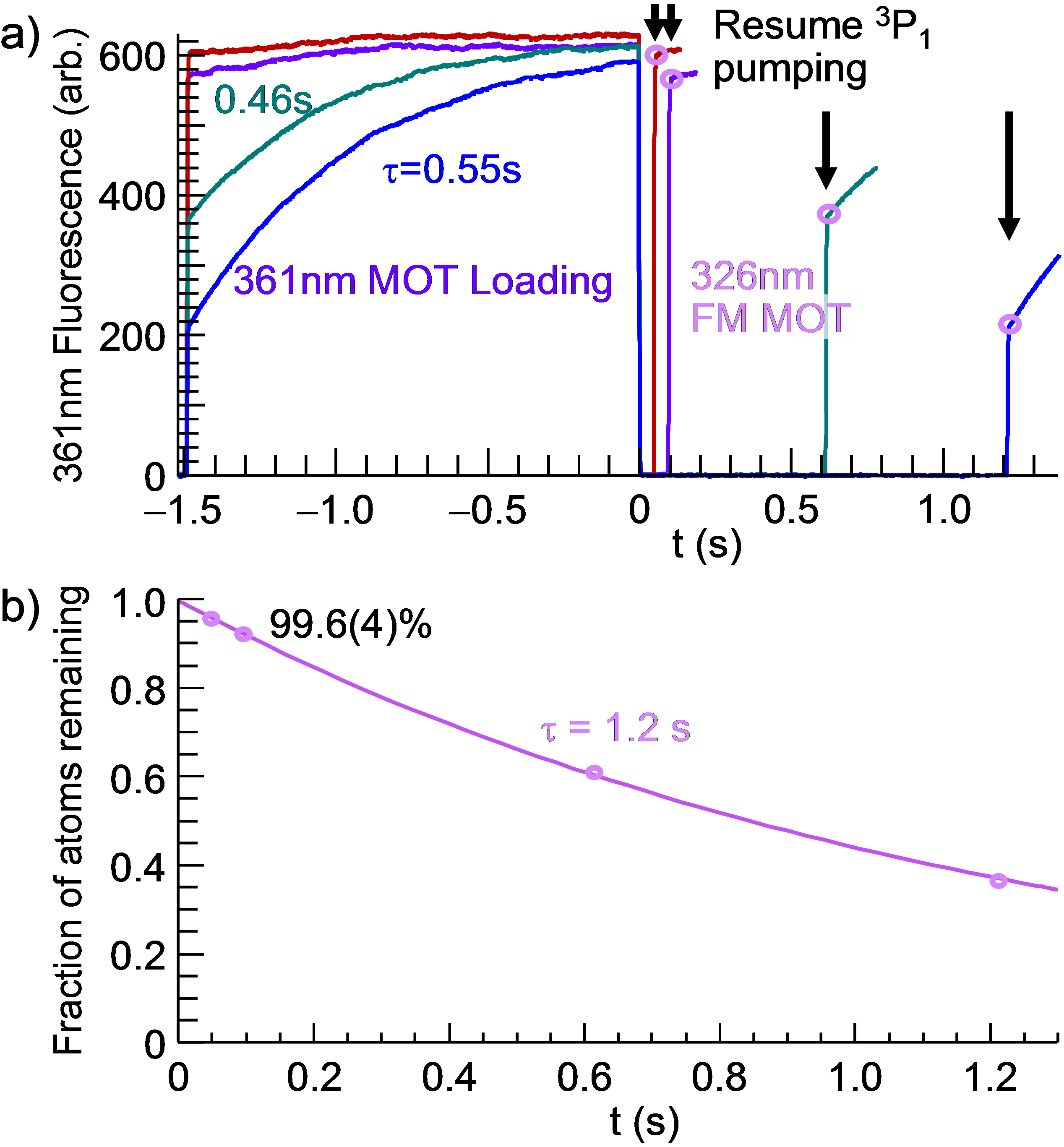}
\caption{\label{fig:480pulse}
a) Inhibiting the $480\,$nm $^3P_1$$\,\rightarrow$$\,^3S_1$$\,\rightarrow$$\,^3P_2$ optical pumping in Fig. \ref{fig:361MOT}a) for a variable time transfers atoms from the metastable MOT to the the narrow-line MOT and back to the metastable MOT. 
b) A fit of the fraction of atoms remaining in the narrow-line MOT versus time shows 99.6(4)\% of the atoms are captured and the lifetime is $1.2\,$s, well longer than the $\approx 0.5\,$s loading time of the metastable MOT. During the narrow-line MOT capture, the magnetic field gradient decreases from $5.7\,$G$\,$cm$^{-1}$ to $0.6\,$G$\,$cm$^{-1}$. 
}
\end{center}
\end{figure}

The low temperatures and small quadrupole magnetic field of the $361\,$nm metastable MOT facilitates an essentially complete transfer of captured atoms to the $326\,$nm intercombination MOT in Fig.~\ref{fig:480pulse}. This compares to 20\% to 30\% direct transfer efficiencies from singlet to narrow-line Ca and Sr MOTs \cite{CaQuenched,LoftusJun2004}. 
Here, at $t=0$, the $480\,$nm $^3P_1$$\,\rightarrow$$\,^3S_1$ optical pumping light is extinguished. The $361\,$nm fluorescence decays in $0.7\,$ms as atoms no longer populate the $^3P_2$ state, e.g., through cascaded spontaneous emission from $^3D_3$$\,\rightarrow\,$5s6p$\,^3P_2$$\,\rightarrow$$\,^3S_1$$\,\rightarrow$$\,^3P_1$$\,\rightarrow$$\,^1S_0$, and may then be captured by the FM-broadened $326\,$nm intercombination MOT. Resuming the $480\,$nm $^3P_1$$\,\rightarrow$$\,^3S_1$$\,\rightarrow$$\,^3P_2$ optical pumping with variable delays shows that $99.6(4)\%$ of the atoms trapped in the metastable MOT are transferred to the intercombination MOT in Fig.~\ref{fig:480pulse}b) and that it has a $1.2\,$s lifetime. Notably, this is significantly longer than the $\approx 0.5\,$s $361\,$nm MOT loading time in Fig.~\ref{fig:480pulse}a) - if background gas collisions limited the lifetimes of both MOTs, the smaller trapping forces of the $326\,$nm MOT would normally yield a shorter lifetime. 

Here, fine structure changing $^3P_2$ collisions, energy-pooling collisions or light-assisted collisions may shorten the loading time in Fig.~\ref{fig:480pulse}a) and the number of atoms in Fig.~\ref{fig:361MOT}b) in the $361\,$nm metastable MOT \cite{Ca3P2Inelastic,Killian3P,EnergyPooling}. In Fig.~\ref{fig:361MOT}b), the initial $361\,$nm MOT loading rates for $^{114}$Cd, $^{111}$Cd and $^{108}$Cd are approximately proportional to their isotopic abundances while the asymptotic numbers of trapped atoms is limited more quickly for the abundant isotopes, in spite of a higher background Cd pressure for the $^{108}$Cd data. Further supporting that $^3P_2$ collisions may limit the number of atoms trapped, the $361\,$nm fluorescence in Fig.~\ref{fig:361MOTscan} (blue curve) abruptly flattens as the metastable MOT light frequency increases beyond $-7\,$MHz of resonance, and then the fluorescence suddenly increases at $-3\,$MHz before falling rapidly to 0 at resonance. From images of the trap, the MOT becomes unstable with large atom numbers at these small detunings, as expected from repulsive radiation pressure forces \cite{Sesko}. Interestingly, here, this MOT instability appears to allow a larger number of trapped atoms with a lower density.

\begin{figure}[t]
\begin{center}
\includegraphics[width=4in]{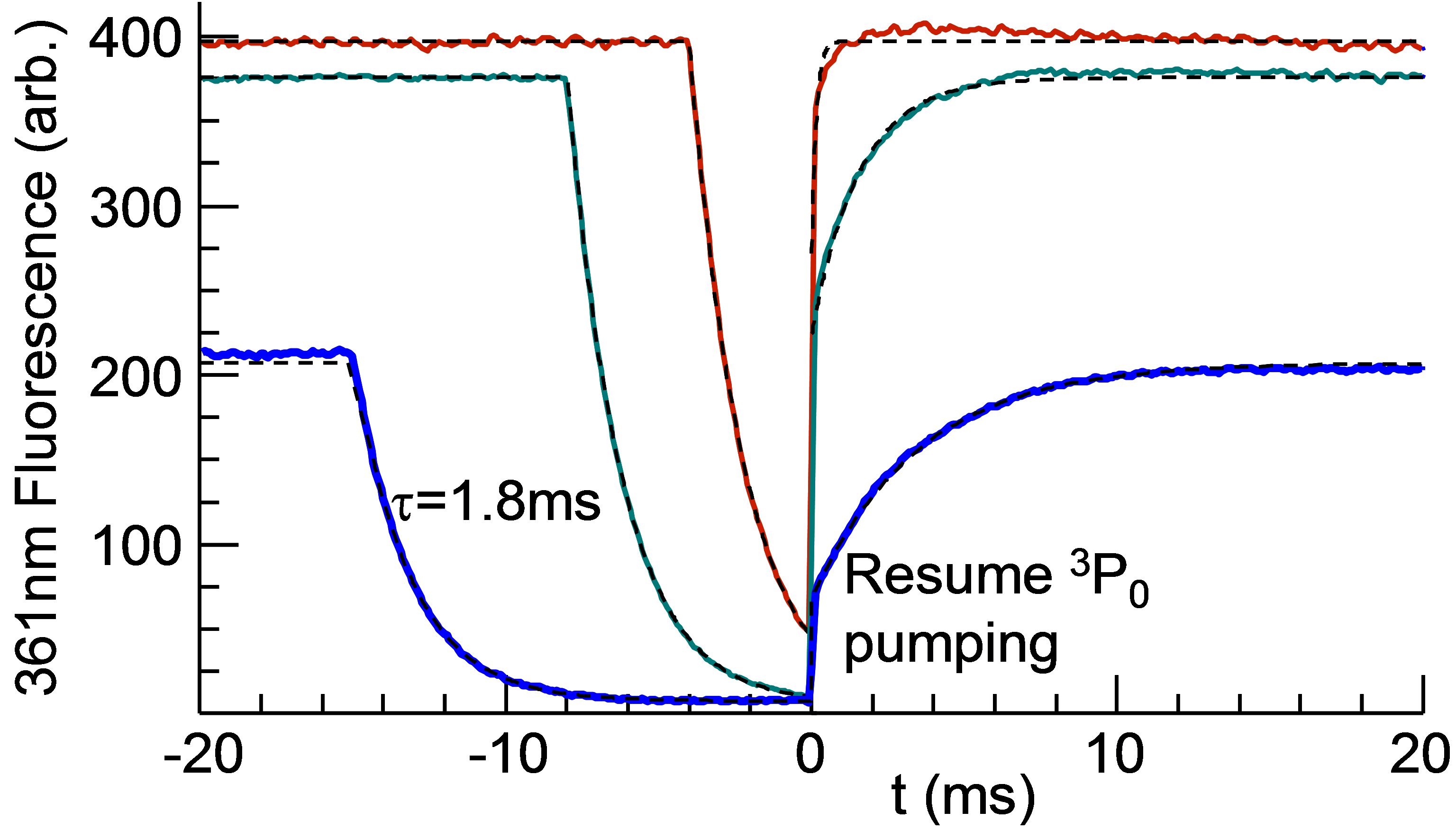}
\caption{\label{fig:468pulse}
Atoms accumulate in the $^3P_0$ excited clock state with a time constant of $1.8\,$ms after the $468\,$nm $^3P_0$$\,\rightarrow$$\,^3S_1$ optical pumping in Fig. \ref{fig:361MOT}a) is inhibited. The black dashed curves are exponential fits to the red, green and blue curves. For the red curve, after $\,^3P_0$ atoms are repumped to $^3P_2$, more atoms are detected at $t=4\,$ms than at $t=-4\,$ms, when the optical pumping is turned off. 
This suggests that $\,^3P_2$ collisional loss limits the number of atoms in the metastable MOT. Here, the cycle repeats every $68\,$ms, which traps fewer atoms for the $14\,$ms $^3P_0$ holding time (blue curve).}
\end{center}
\end{figure}

Figure~\ref{fig:468pulse} shows the $361\,$nm metastable MOT fluorescence when the $468\,$nm $^3P_1$$\,\rightarrow$$\,^3S_1$ optical pumping light is extinguished for $\tau_e = 4\,$ms to $14\,$ms. This transfers cold $^3P_2$ atoms to $^3P_0$, e.g., via decay from $^3D_3$ to 5s6p$\,^3P_2$ in Fig.~\ref{fig:361MOT}a). 
These $^3P_0$ atoms are not trapped and begin to slowly leave the detection region, yielding less fluorescence just after they are repumped to $^3P_2$ at $t=0$. 
Because the atoms are recaptured in much less than the MOT loading time, they do not exit the MOT laser beams volume during this time.
After the $468\,$nm light is extinguished, the $326\,$nm excitation to $^3P_1$ and the $480\,$nm pumping from $^3P_1$ to $^3S_1$ to $^3P_2$, continues to load the $361\,$nm metastable MOT, albeit at a lower rate because some $^3P_1$ atoms excited to $^3S_1$ decay to $^3P_0$ instead of $^3P_2$. 
Nonetheless, for $\tau_e = 4\,$ms, the $361\,$nm MOT fluorescence at $t=4\,$ms is greater than at $t=-4\,$ms, consistent with a lower loss rate for atoms in $^3P_0$ than in $^3P_2$, and the dominant $^3P_2$ loss being fine structure changing collisions. 
For $\tau_e = 14\,$ms in Fig.~\ref{fig:468pulse} (blue curve), the smaller number of trapped atoms is due to the short cycle time of $68\,$ms.

\section{Laser System}
To generate the four laser sources in Fig.~\ref{fig:361MOT}a)\cite{OSA, CdISS, MCFS}, we begin with a $1083\,$nm extended-cavity diode laser (ECDL) that seeds a fiber amplifier, which is frequency doubled to $542\,$nm with LBO in a resonant cavity. We generate $361\,$nm UVA light for the metastable MOT with doubly resonant sum-frequency generation (SFG) of the $542\,$nm and $1083\,$nm light. 
The $326\,$nm UVA light for the intercombination MOT \cite{OSA, CdISS, MCFS,CdMag,Poli326} is made in a second doubly resonant BBO cavity for SFG of the $542\,$nm light and $820\,$nm light from an ECDL and tapered amplifier. 
Finally, the low power blue $468\,$nm and $480\,$nm optical pumping light is generated with single-pass SFG of $1083\,$nm with $823\,$nm and $862\,$nm light in periodically-poled lithium niobate (PPLN) waveguides. 
The $1083\,$nm ECDL is frequency stabilised with a tunable offset to a ring reference-cavity, which is also used to monitor the $820\,$nm, $823\,$nm and $862\,$nm ECDL frequencies \cite{CdISS}. Saturated absorption of molecular I\textsubscript{2} provides an absolute frequency reference at $542\,$nm. 

\section{Isotope Shift and Absolute Frequency Measurements}
We use the sharp blue-edge of the MOT fluorescence in Fig.~\ref{fig:361MOTscan} to measure the isotope shifts \cite{CdISS} and the absolute frequency of the $^3P_2$$\,\rightarrow$ $\,^3D_3$ transition. 
Relative to the the R(83)28-0 I\textsubscript{2} a\textsubscript{7} hyperfine component, the $^{114}$Cd $^3P_2$$\,\rightarrow$$\,^3D_3$ transition frequency is $\nu_{114}=\frac{3}{2}\nu_{R(83)28-0\,a_7}+261(3)\,$MHz, where a model of a number of I\textsubscript{2} absolute frequencies gives $\nu_{R(83)28-0\,a_7}=553\ 397\ 541(10)\,$MHz \cite{TiemannPrivComm, TiemannI2}. 
This improves the previous uncertainty \cite{Burns56} by a factor of 20, yielding a 5s5d$\,^3D_3$ energy of $59\ 515.990\,$cm$^{-1}$ for natural Cd, the value of \cite{Burns56}, after a likely misprint is corrected \cite{KramidaPrivComm}. 
The previous $300\,$MHz uncertainty contributed to some difficultly initially finding this broad transition - changing the frequency of our $361\,$nm light by more than the 10 MHz tuning range of the AOM requires changing the $820\,$nm laser frequency and the dispersion of the $326\,$nm doubly-resonant SFG cavity to preserve the narrow-line MOT \cite{OSA}. 
We use the blue MOT edge to measure the isotope shifts for the Cd bosons \cite{CdISS} and fermions, which are given in Table \ref{tab:ISS}. We similarly report the $480\,$nm $^3P_1$$\,\rightarrow$$\,^3S_1$ ISs from the optical pumping rate of the metastable MOT fluorescence, as in \cite{CdISS}, which reported these ISs for Cd bosons.

\section{Conclusions}
Cadmium has a number of attractive properties for atomic clocks and interferometers and quantum gas experiments.
These include narrow linewidth clock and other intercombination transitions, a small sensitivity of its clock transition to BBR, small sensitivities to external magnetic fields, a level structure that requires minimal repumping for laser cooling, and six bosonic isotopes and two nuclear-spin-½ fermionic isotopes. We demonstrate that all 8 Cd isotopes can be efficiently laser-cooled without exciting the hard UVC $229\,$nm $^1S_0$$\,\rightarrow$$\,^1P_1$ transition. We begin by loading a MOT using only the Cd $67\,$kHz wide $326\,$nm $^1S_0$$\,\rightarrow$ $\,^3P_1$ UVA intercombination transition, the narrowest transition used to capture room temperature atoms. We dramatically increase the laser-cooling capture rate using the dipole-allowed $361\,$nm $^3P_2$$\,\rightarrow$$\,^3D_3$ UVA transition to load a metastable MOT, with no additional repumping required for the Cd fermions.
The $J=2$ or $F=\frac{5}{2}$ substate structure of $^3P_2$ lower level of the metastable Cd MOT, along with the favorable $^3P_2$$\,\rightarrow$$\,^3D_3$ hyperfine structure, yields the usual MOT behaviors of alkali atoms, in contrast to those of alkaline-earth singlet and intercombination MOTs. 
The $^3P_2$$\,\rightarrow$$\,^3D_3$ metastable Cd MOT avoids cumbersomely large magnetic field gradients, provides low temperatures, and enables a 100\% transfer of atoms to the narrow-line $326\,$nm MOT for subsequent cooling to microkelvin temperatures \cite{CdMag}. 
We observe evidence of collisional loss in the metastable MOT, report an absolute frequency of the $^3P_2$$\,\rightarrow$$\,^3D_3$ transition, reducing its uncertainty by a factor of 20, as well as its isotope shifts and those of the $326\,$nm and $480\,$nm transitions \cite{CdISS} for the $^{111}$Cd and $^{113}$Cd fermions. 

Capturing atoms with $^3P_2$$\,\rightarrow$$\,^3D_3$ metastable MOTs may also efficiently collect atoms from an atomic beam that is optically pumped to $^3P_2$ and slowed with the $^3P_2$$\,\rightarrow$$\,^3D_3$ transition, or a beam slowed by low-power $^1S_0$$\,\rightarrow$$\,^1P_1$ light, avoiding the $^1S_0$ $\,\rightarrow$ $\,^1P_1$ MOT and its required large magnetic field gradients \cite{Vladen_3P1only_and1P12015, YbSingletslowerTripletMOT, YbCoreMOT}. 
Beyond Cd, these techniques may be particularly useful for Hg \cite{BizeMore254,Stellmer_Hg_highphasespace2022}, requiring only low power UVC light, as well as for Sr \cite{KatoriSr3P2Cool} and Yb, and possibly Zn \cite{StellmerZn}, Ca and Mg \cite{Rasel3D3Det}. Avoiding UVC light, or reducing its required power, is advantageous for long-term operation of future Zn, Cd and Hg clocks and other quantum sensors, particularly in remote environments.

\section*{Acknowledgments}
We gratefully acknowledge stimulating  and helpful conversations with U. Sterr, S. Gupta, A. Yamaguchi, H. Katori, A. Kramida, and E. Tiemmann and contributions of D. Schussheim \cite{OSA,MCFS}.
This material is based upon work supported by the U.S. National Science Foundation under award No. 2012117.

\section*{References}


\begin{thebibliography}{10}
\expandafter\ifx\csname url\endcsname\relax
  \def\url#1{{\tt #1}}\fi
\expandafter\ifx\csname urlprefix\endcsname\relax\def\urlprefix{}\fi
\providecommand{\eprint}[2][]{\url{#2}}

\bibitem{ClockRevRMP15}
Ludlow A~D, Boyd M~M, Ye J, Peik E and Schmidt P~O 2015 {\em Rev. Mod. Phys.\/} {\bf 87}(2) 637--701 \urlprefix\url{https://link.aps.org/doi/10.1103/RevModPhys.87.637}

\bibitem{BACON2021_AlSrYb}
Beloy K~{\it et al} 2021 {\em Nature\/} {\bf 591} 564--569 ISSN 1476-4687 \urlprefix\url{https://doi.org/10.1038/s41586-021-03253-4}

\bibitem{BizeMore254}
McFerran J~J, Yi L, Mejri S, Di~Manno S, Zhang W, Gu\'ena J, Le~Coq Y and Bize S 2012 {\em Phys. Rev. Lett.\/} {\bf 108}(18) 183004 \urlprefix\url{https://link.aps.org/doi/10.1103/PhysRevLett.108.183004}

\bibitem{Ohmae2020_HgYbSrloop}
Ohmae N, Bregolin F, Nemitz N and Katori H 2020 {\em Opt. Express\/} {\bf 28} 15112--15121 \urlprefix\url{http://www.opticsexpress.org/abstract.cfm?URI=oe-28-10-15112}

\bibitem{TinoSr8788EEP2014}
Tarallo M~G, Mazzoni T, Poli N, Sutyrin D~V, Zhang X and Tino G~M 2014 {\em Phys. Rev. Lett.\/} {\bf 113}(2) 023005 \urlprefix\url{https://link.aps.org/doi/10.1103/PhysRevLett.113.023005}

\bibitem{MAGIS100_2021}
Abe M~{\it et al} 2021 {\em Quantum Science and Technology\/} {\bf 6} 044003 \urlprefix\url{https://dx.doi.org/10.1088/2058-9565/abf719}

\bibitem{Tino_AIrev_2021}
Tino G~M 2021 {\em Quantum Science and Technology\/} {\bf 6} 024014 \urlprefix\url{https://dx.doi.org/10.1088/2058-9565/abd83e}

\bibitem{Poli_CdSr_Source2023}
Bandarupally S, Tinsley J~N, Chiarotti M and Poli N 2023 {\em Journal of Physics B: Atomic, Molecular and Optical Physics\/} {\bf 56} 185301 \urlprefix\url{https://dx.doi.org/10.1088/1361-6455/acf3bf}

\bibitem{Takahashi_YbBEC2003}
Takasu Y, Maki K, Komori K, Takano T, Honda K, Kumakura M, Yabuzaki T and Takahashi Y 2003 {\em Phys. Rev. Lett.\/} {\bf 91}(4) 040404 \urlprefix\url{https://link.aps.org/doi/10.1103/PhysRevLett.91.040404}

\bibitem{PTB_Ca_BEC2009}
Kraft S, Vogt F, Appel O, Riehle F and Sterr U 2009 {\em Phys. Rev. Lett.\/} {\bf 103}(13) 130401 \urlprefix\url{https://link.aps.org/doi/10.1103/PhysRevLett.103.130401}

\bibitem{Schreck_Sr84BEC_2009}
Stellmer S, Tey M~K, Huang B, Grimm R and Schreck F 2009 {\em Phys. Rev. Lett.\/} {\bf 103}(20) 200401 \urlprefix\url{https://link.aps.org/doi/10.1103/PhysRevLett.103.200401}

\bibitem{Killian_Sr84BEC_2009}
de~Escobar Y~N~M, Mickelson P~G, Yan M, DeSalvo B~J, Nagel S~B and Killian T~C 2009 {\em Phys. Rev. Lett.\/} {\bf 103}(20) 200402 \urlprefix\url{https://link.aps.org/doi/10.1103/PhysRevLett.103.200402}

\bibitem{Rey_Fermigases_SU(N)_2014}
Cazalilla M~A and Rey A~M 2014 {\em Reports on Progress in Physics\/} {\bf 77} 124401 \urlprefix\url{https://dx.doi.org/10.1088/0034-4885/77/12/124401}

\bibitem{Chen2022CWBEC}
Chen C~C, Gonz{\'a}lez~Escudero R, Min{\'a}{\v{r}} J, Pasquiou B, Bennetts S and Schreck F 2022 {\em Nature\/} {\bf 606} 683--687 ISSN 1476-4687 \urlprefix\url{https://www.nature.com/articles/s41586-022-04731-z}

\bibitem{SU_N_Group2_2010}
Gorshkov A~V, Hermele M, Gurarie V, Xu C, Julienne P~S, Ye J, Zoller P, Demler E, Lukin M~D and Rey A~M 2010 {\em Nature Physics\/} {\bf 6} 289--295 ISSN 1745-2481 \urlprefix\url{https://doi.org/10.1038/nphys1535}

\bibitem{Daley_Group2_QC_QI_2011}
Daley A~J 2011 {\em Quantum Information Processing\/} {\bf 10} 865 ISSN 1573-1332 \urlprefix\url{https://doi.org/10.1007/s11128-011-0293-3}

\bibitem{Foss_Feig_QI_2019}
Pagano G, Scazza F and Foss-Feig M 2019 {\em Advanced Quantum Technologies\/} {\bf 2} 1800067 \urlprefix\url{https://onlinelibrary.wiley.com/doi/abs/10.1002/qute.201800067}

\bibitem{AE_Rydberg_Gate2020}
Madjarov I~S, Covey J~P, Shaw A~L, Choi J, Kale A, Cooper A, Pichler H, Schkolnik V, Williams J~R and Endres M 2020 {\em Nature Physics\/} {\bf 16} 857--861 ISSN 1745-2481 \urlprefix\url{https://doi.org/10.1038/s41567-020-0903-z}

\bibitem{Bloch2012QSim}
Bloch I, Dalibard J and Nascimb{\`e}ne S 2012 {\em Nat. Phys.\/} {\bf 8} 267--276 ISSN 1745-2481 \urlprefix\url{https://doi.org/10.1038/nphys2259}

\bibitem{QSim_Takahashi2020}
Sch{\"a}fer F, Fukuhara T, Sugawa S, Takasu Y and Takahashi Y 2020 {\em Nature Reviews Physics\/} {\bf 2} 411--425 ISSN 2522-5820 \urlprefix\url{https://doi.org/10.1038/s42254-020-0195-3}

\bibitem{2020-Yb}
Counts I, Hur J, Aude~Craik D~P~L, Jeon H, Leung C, Berengut J~C, Geddes A, Kawasaki A, Jhe W and Vuleti\ifmmode~\acute{c}\else \'{c}\fi{} V 2020 {\em Phys. Rev. Lett.\/} {\bf 125}(12) 123002 \urlprefix\url{https://link.aps.org/doi/10.1103/PhysRevLett.125.123002}

\bibitem{2021-Yb}
Ono K, Saito Y, Ishiyama T, Higomoto T, Takano T, Takasu Y, Yamamoto Y, Tanaka M and Takahashi Y 2022 {\em Phys. Rev. X\/} {\bf 12}(2) 021033 \urlprefix\url{https://link.aps.org/doi/10.1103/PhysRevX.12.021033}

\bibitem{2021-Yb2}
Figueroa N~L, Berengut J~C, Dzuba V~A, Flambaum V~V, Budker D and Antypas D 2022 {\em Phys. Rev. Lett.\/} {\bf 128}(7) 073001 \urlprefix\url{https://link.aps.org/doi/10.1103/PhysRevLett.128.073001}

\bibitem{2022-Yb}
Hur J~{\it et al} 2022 {\em Phys. Rev. Lett.\/} {\bf 128}(16) 163201 \urlprefix\url{https://link.aps.org/doi/10.1103/PhysRevLett.128.163201}

\bibitem{CaISS}
Solaro C, Meyer S, Fisher K, Berengut J~C, Fuchs E and Drewsen M 2020 {\em Phys. Rev. Lett.\/} {\bf 125}(12) 123003 \urlprefix\url{https://link.aps.org/doi/10.1103/PhysRevLett.125.123003}

\bibitem{KolkowitzSrIS}
Zheng X, Dolde J, Lochab V, Merriman B~N, Li H and Kolkowitz S 2022 {\em Nature\/} {\bf 602} 425--430 \urlprefix\url{https://doi.org/10.1038/s41467-023-40629-8}

\bibitem{CdISS}
Ohayon B, Hofs\"ass S, Padilla-Castillo J~E, Wright S~C, Meijer G, Truppe S, Gibble K and Sahoo B~K 2022 {\em New Journal of Physics\/} {\bf 24} 123040 \urlprefix\url{https://dx.doi.org/10.1088/1367-2630/acacbb}

\bibitem{CdMag}
Yamaguchi A, Safronova M~S, Gibble K and Katori H 2019 {\em Phys. Rev. Lett.\/} {\bf 123}(11) 113201 \urlprefix\url{https://link.aps.org/doi/10.1103/PhysRevLett.123.113201}

\bibitem{Derevianko_ZnCdBBR_2019}
Dzuba V~A and Derevianko A 2019 {\em Journal of Physics B: Atomic, Molecular and Optical Physics\/} {\bf 52} 215005 \urlprefix\url{https://dx.doi.org/10.1088/1361-6455/ab4434}

\bibitem{MonroeCd}
Brickman K~A, Chang M~S, Acton M, Chew A, Matsukevich D, Haljan P~C, Bagnato V~S and Monroe C 2007 {\em Phys. Rev. A\/} {\bf 76}(4) 043411 \urlprefix\url{https://link.aps.org/doi/10.1103/PhysRevA.76.043411}

\bibitem{Katori229}
Kaneda Y, Yarborough J~M, Merzlyak Y, Yamaguchi A, Hayashida K, Ohmae N and Katori H 2016 {\em Opt. Lett.\/} {\bf 41} 705--708 \urlprefix\url{https://opg.optica.org/ol/abstract.cfm?URI=ol-41-4-705}

\bibitem{Stellmer_Hg_highphasespace2022}
Lavigne Q, Groh T and Stellmer S 2022 {\em Phys. Rev. A\/} {\bf 105}(3) 033106 \urlprefix\url{https://link.aps.org/doi/10.1103/PhysRevA.105.033106}

\bibitem{Poli229}
Tinsley J~N, Bandarupally S, Penttinen J~P, Manzoor S, Ranta S, Salvi L, Guina M and Poli N 2021 {\em Opt. Express\/} {\bf 29} 25462--25476 \urlprefix\url{https://opg.optica.org/oe/abstract.cfm?URI=oe-29-16-25462}

\bibitem{TruppeFermionISS}
Hofs\"ass S, Padilla-Castillo J~E, Wright S~C, Kray S, Thomas R, Sartakov B~G, Ohayon B, Meijer G and Truppe S 2023 {\em Phys. Rev. Res.\/} {\bf 5}(1) 013043 \urlprefix\url{https://link.aps.org/doi/10.1103/PhysRevResearch.5.013043}

\bibitem{Cd3P1Hyperfine1964}
Lurio A and Novick R 1964 {\em Phys. Rev.\/} {\bf 134}(3A) A608--A614 \urlprefix\url{https://link.aps.org/doi/10.1103/PhysRev.134.A608}

\bibitem{Cd3p1Lifetime1989}
Czajkowski M, Bobkowski R and Krause L 1989 {\em Phys. Rev. A\/} {\bf 40}(8) 4338--4343 , this more recent measurement suggests a $3.0(1)\,\mu$s lifetime and a $53(2)\,$kHz linewidth \urlprefix\url{https://link.aps.org/doi/10.1103/PhysRevA.40.4338}

\bibitem{ThompsonNarrowStimCooling}
Norcia M~A, Cline J~R~K, Bartolotta J~P, Holland M~J and Thompson J~K 2018 {\em New Journal of Physics\/} {\bf 20} 023021 \urlprefix\url{https://dx.doi.org/10.1088/1367-2630/aaa950}

\bibitem{Vladen_3P1only_and1P12015}
Kawasaki A, Braverman B, Yu Q and Vuletic V 2015 {\em Journal of Physics B: Atomic, Molecular and Optical Physics\/} {\bf 48} 155302 , report capturing Yb using only its $184\,$kHz $556\,$nm intercombination transition. \urlprefix\url{https://dx.doi.org/10.1088/0953-4075/48/15/155302}

\bibitem{YbSingletslowerTripletMOT}
Kuwamoto T, Honda K, Takahashi Y and Yabuzaki T 1999 {\em Phys. Rev. A\/} {\bf 60}(2) R745--R748 \urlprefix\url{https://link.aps.org/doi/10.1103/PhysRevA.60.R745}

\bibitem{YbCoreMOT}
Lee J, Lee J~H, Noh J and Mun J 2015 {\em Phys. Rev. A\/} {\bf 91}(5) 053405 \urlprefix\url{https://link.aps.org/doi/10.1103/PhysRevA.91.053405}

\bibitem{Ca3p23D3MOT}
Gr\"unert J and Hemmerich A 2002 {\em Phys. Rev. A\/} {\bf 65}(4) 041401 \urlprefix\url{https://link.aps.org/doi/10.1103/PhysRevA.65.041401}

\bibitem{Rasel3D3Det}
Friebe J~{\it et al} 2011 {\em New Journal of Physics\/} {\bf 13} 125010 \urlprefix\url{https://dx.doi.org/10.1088/1367-2630/13/12/125010}

\bibitem{KatoriSr3P2Cool}
Akatsuka T, Hashiguchi K, Takahashi T, Ohmae N, Takamoto M and Katori H 2021 {\em Phys. Rev. A\/} {\bf 103}(2) 023331 \urlprefix\url{https://link.aps.org/doi/10.1103/PhysRevA.103.023331}

\bibitem{LudlowDarkCool}
Zhang X, Beloy K, Hassan Y~S, McGrew W~F, Chen C~C, Siegel J~L, Grogan T and Ludlow A~D 2022 {\em Phys. Rev. Lett.\/} {\bf 129}(11) 113202 \urlprefix\url{https://link.aps.org/doi/10.1103/PhysRevLett.129.113202}

\bibitem{Ludlow3D1Latt}
Chen C~C, Siegel J~L, Hunt B~D, Grogan T, Hassan Y~S, Beloy K, Gibble K, Brown R~C and Ludlow A~D 2024 {\em Phys. Rev. Lett.\/} {\bf 133}(5) 053401 \urlprefix\url{https://link.aps.org/doi/10.1103/PhysRevLett.133.053401}

\bibitem{GretchenCool3P2}
Barker D~S, Reschovsky B~J, Pisenti N~C and Campbell G~K 2015 {\em Phys. Rev. A\/} {\bf 92}(4) 043418 \urlprefix\url{https://link.aps.org/doi/10.1103/PhysRevA.92.043418}

\bibitem{CaQuenched}
Curtis E~A, Oates C~W and Hollberg L 2001 {\em Phys. Rev. A\/} {\bf 64}(3) 031403 \urlprefix\url{https://link.aps.org/doi/10.1103/PhysRevA.64.031403}

\bibitem{PTBNISTCa}
Sterr U, Degenhardt C, Stoehr H, Lisdat C, Schnatz H, Helmcke J, Riehle F, Wilpers G, Oates C and Hollberg L 2004 {\em Comptes Rendus Physique\/} {\bf 5} 845--855 ISSN 1631-0705 fundamental metrology \urlprefix\url{https://www.sciencedirect.com/science/article/pii/S1631070504001525}

\bibitem{Cd3S1HyperfineEPJ}
Fr{\"o}mmgen N~{\it et al} 2015 {\em The European Physical Journal D\/} {\bf 69} 164 ISSN 1434-6079 \urlprefix\url{https://doi.org/10.1140/epjd/e2015-60219-0}

\bibitem{Cd3S1HyperfineRoyal}
Brimicombe M~S~W~M, Stacey D~N, Stacey V, H\"uhnermann H, Menzel N and Bleaney B 1976 {\em Proceedings of the Royal Society of London. A. Mathematical and Physical Sciences\/} {\bf 352} 141--152 \urlprefix\url{https://royalsocietypublishing.org/doi/abs/10.1098/rspa.1976.0168}

\bibitem{LichtenCd3P2Hyperfine}
Faust W, McDermott M and Lichten W 1960 {\em Phys. Rev.\/} {\bf 120}(2) 469--469 \urlprefix\url{https://link.aps.org/doi/10.1103/PhysRev.120.469}

\bibitem{Cd3DHyperfine}
Chantepie, M and Lecluse, Y 1971 {\em J. Phys. France\/} {\bf 32} 415--419 \urlprefix\url{https://doi.org/10.1051/jphys:01971003205-6041500}

\bibitem{Cd326Details}
The MOT is loaded for $400\,$ms and the atom fluorescence is integrated for $16.67\,$ms after the laser's FM is inhibited and its intensity and detuning are ramped \cite{MCFS}. A subsequent background, also integrated for $16.67\,$ms after the atoms are cleared with a blue laser detuning, is subtracted. The direction of the MOT magnetic field gradient is then reversed and its background, from two more integrations, is additionally subtracted, giving an updated number of trapped atoms every $907\,$ms.

\bibitem{RolstenBlue}
Walhout M, Megens H~J~L, Witte A and Rolston S~L 1993 {\em Phys. Rev. A\/} {\bf 48}(2) R879--R882 \urlprefix\url{https://link.aps.org/doi/10.1103/PhysRevA.48.R879}

\bibitem{MonsterMOT}
Gibble K~E, Kasapi S and Chu S 1992 {\em Opt. Lett.\/} {\bf 17} 526--528 \urlprefix\url{https://opg.optica.org/ol/abstract.cfm?URI=ol-17-7-526}

\bibitem{LoftusJun2004}
Loftus T~H, Ido T, Ludlow A~D, Boyd M~M and Ye J 2004 {\em Phys. Rev. Lett.\/} {\bf 93}(7) 073003 \urlprefix\url{https://link.aps.org/doi/10.1103/PhysRevLett.93.073003}

\bibitem{Ca3P2Inelastic}
Hansen D and Hemmerich A 2006 {\em Phys. Rev. Lett.\/} {\bf 96}(7) 073003 \urlprefix\url{https://link.aps.org/doi/10.1103/PhysRevLett.96.073003}

\bibitem{Killian3P}
Traverso A, Chakraborty R, Martinez~de Escobar Y~N, Mickelson P~G, Nagel S~B, Yan M and Killian T~C 2009 {\em Phys. Rev. A\/} {\bf 79}(6) 060702 \urlprefix\url{https://link.aps.org/doi/10.1103/PhysRevA.79.060702}

\bibitem{EnergyPooling}
Kelly J~F, Harris M and Gallagher A 1988 {\em Phys. Rev. A\/} {\bf 38}(3) 1225--1229 \urlprefix\url{https://link.aps.org/doi/10.1103/PhysRevA.38.1225}

\bibitem{Sesko}
Walker T, Sesko D and Wieman C 1990 {\em Phys. Rev. Lett.\/} {\bf 64}(4) 408--411 \urlprefix\url{https://link.aps.org/doi/10.1103/PhysRevLett.64.408}

\bibitem{OSA}
Schussheim D and Gibble K 2018 Laser system to laser-cool and trap cadmium: towards a cadmium optical lattice clock {\em Frontiers in Optics / Laser Science\/} (Optical Society of America) p LTh1F.2 \urlprefix\url{http://www.osapublishing.org/abstract.cfm?URI=LS-2018-LTh1F.2}

\bibitem{MCFS}
Schussheim D~T and Gibble K 2023 {\em Review of Scientific Instruments\/} {\bf 94} 085101 ISSN 0034-6748 \urlprefix\url{https://doi.org/10.1063/5.0157330}

\bibitem{Poli326}
Manzoor S, Tinsley J~N, Bandarupally S, Chiarotti M and Poli N 2022 {\em Opt. Lett.\/} {\bf 47} 2582--2585 \urlprefix\url{https://opg.optica.org/ol/abstract.cfm?URI=ol-47-10-2582}

\bibitem{TiemannPrivComm}
Tiemann E 2022 {Private Communication}

\bibitem{TiemannI2}
Bodermann B, Kn\"{o}ckel H and Tiemann E 2002 {\em The European Physical Journal D\/} {\bf 19} 31--44 \urlprefix\url{https://doi.org/10.1140/epjd/e20020052}

\bibitem{Burns56}
Burns K and Adams K B 1956 \textit{J. Opt. Soc. Am.} \textbf{46} 94--99. From their measured wavelength, it is likely that their calculated $\mathrm{5s5d}\,^3\mathrm{D}_3$ energy was $59\ 515.990\,$cm$^{-1}$ \cite{KramidaPrivComm}. \urlprefix\url{https://opg.optica.org/abstract.cfm?URI=josa-46-2-94}

\bibitem{KramidaPrivComm}
Kramida A 2022 {Private Communication}

\bibitem{StellmerZn}
B\"{u}ki M, R\"{o}ser D and Stellmer S 2021 {\em Appl. Opt.\/} {\bf 60} 9915--9918 \urlprefix\url{https://opg.optica.org/ao/abstract.cfm?URI=ao-60-31-9915}

\end{thebibliography}

\providecommand{\newblock}{}

\end{document}